\documentclass[11pt,a4paper]{article}

\usepackage[T1]{fontenc}
\usepackage[utf8]{inputenc}
\usepackage{libertinus} 
\usepackage{microtype}
\usepackage{geometry}
\usepackage{graphicx}
\usepackage{booktabs}
\usepackage{enumitem}
\usepackage{xcolor}
\usepackage{comment} 
\usepackage{float}
\usepackage{mdframed}
\newmdenv[linecolor=blue,backgroundcolor=blue!10,roundcorner=5pt]{definition}
\newmdenv[linecolor=red,backgroundcolor=red!10,roundcorner=5pt]{warning}
\usepackage[normalem]{ulem}
\usepackage{eso-pic}
\usepackage{verbatim} 
\usepackage{csquotes}
\usepackage[french]{babel}
\addto\captionsfrench{}
\usepackage{listings}
\usepackage{hyperref}
\usepackage{dtk-logos}

\definecolor{codebg}{RGB}{248,248,248}
\definecolor{codeframe}{RGB}{200,200,200}
\definecolor{codekw}{RGB}{0,0,150}
\definecolor{codestring}{RGB}{163,21,21}
\definecolor{codecomment}{RGB}{0,128,0}
\definecolor{codeid}{RGB}{0,0,0}

\lstdefinestyle{accadil}{
  backgroundcolor=\color{codebg},
  frame=single,
  rulecolor=\color{codeframe},
  basicstyle=\ttfamily\small,
  keywordstyle=\color{codekw}\bfseries,
  stringstyle=\color{codestring},
  commentstyle=\color{codecomment}\itshape,
  identifierstyle=\color{codeid},
  breaklines=true,
  columns=fullflexible,
  showstringspaces=false,
  tabsize=2,
  keepspaces=true,
  captionpos=b
}

\lstdefinelanguage{BibTeX}{
  morecomment=[l]{@},
  morecomment=[s]{\{}{\}},
  morestring=[b]"
}

\hypersetup{
  colorlinks=true,
  linkcolor=black,
  urlcolor=blue,
  citecolor=black
}

\title{
\Huge \textbf{
\textsc{Interface Homme-Machine\\
pour l'Identification\\
des Liaisons de Coins}}\\[1em] 
\large {\textbf{Projet ACCADIL}}\\
\textsc{\small (\textbf{A}ncient \textbf{C}oin \textbf{C}lassification \textbf{A}lgorithms for \textbf{D}Ie \textbf{L}inks)}\\[1em]  
\textbf{\Large Guide utilisateur - V1.0} 
}

\author{
\textbf{Patrice Labedan$^{*}$, Nicolas Drougard$^{*}$}\\
\textit{ISAE-SUPAERO, Université de Toulouse, France}\\
$^{*}$\texttt{\{prenom.nom\}@isae-supaero.fr}\\[0.5em]
Avec la collaboration de Francis Dieulafait$^{**}$\\
\textit{HADES, L'Union, France}\\
$^{**}$\texttt{\{prenom.nom\}@hades-archeologie.com}
}
\date{\today}

\begin{document}

\AddToShipoutPictureBG*{%
  \put(0,0){%
    \parbox[b][\paperheight]{\paperwidth}{%
      \vfill
      \centering
      \includegraphics[width=\paperwidth,height=\paperheight,keepaspectratio]{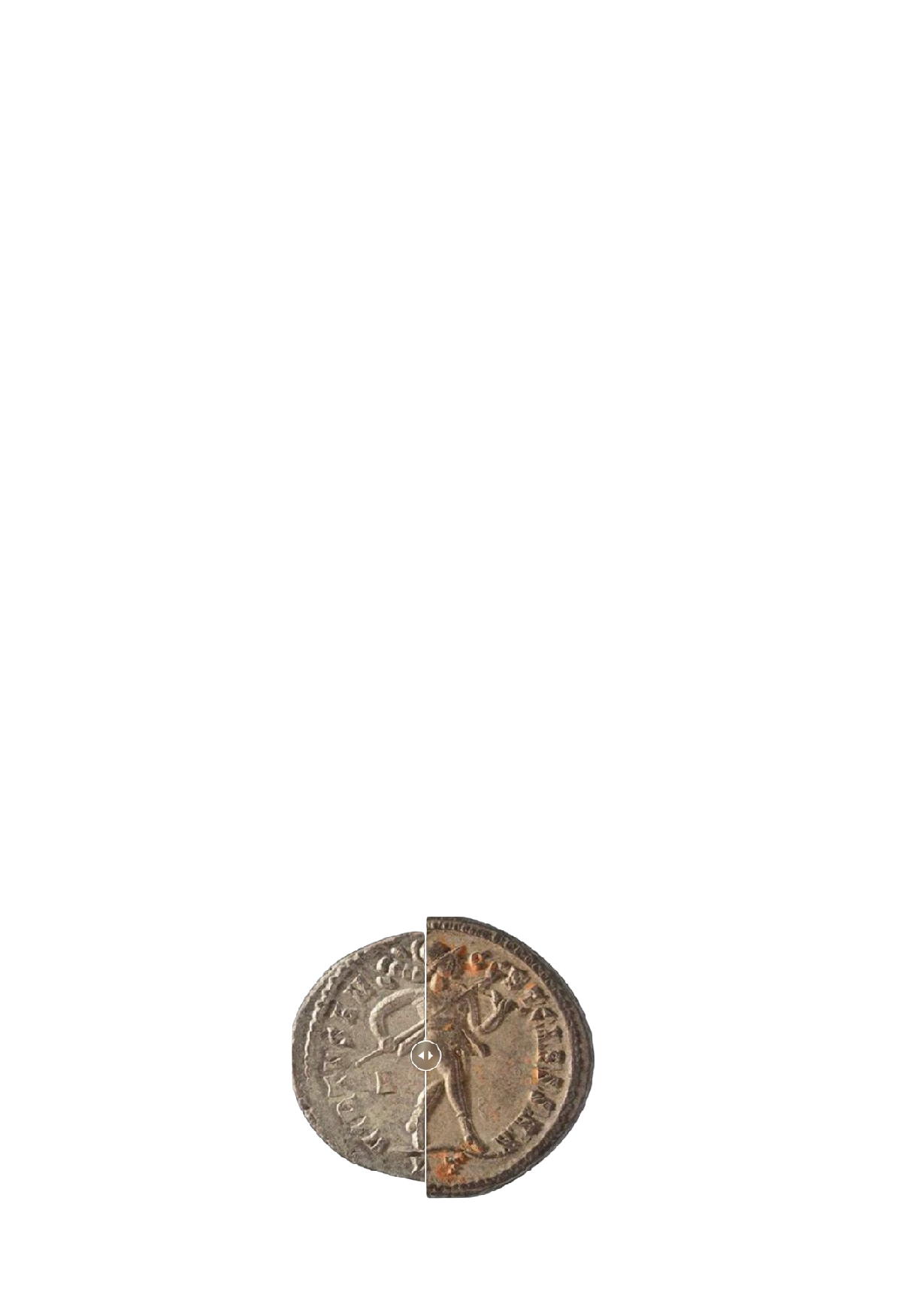}
      \vfill
    }
  }
}

\maketitle

\selectlanguage{french}
\begin{abstract}

ACCADIL est un projet qui a mené au développement d'outils informatiques pour
l’identification des liaisons de coins monétaires à partir de photographies de monnaies. 
Il propose un algorithme de calcul basé sur des techniques de vision par ordinateur et de classification, et une interface en ligne permettant la vérification interactive des résultats. 
Ce guide décrit succinctement les principes algorithmiques, la préparation des données avant l'analyse, ainsi que les fonctionnalités offertes par l’interface~: ajout de jeux de données, modes de visualisation (superposition, côte à côte, loupe, transparence), export des résultats et visualisation des distances. ACCADIL propose ainsi aux numismates un outil complet pour l’analyse des liaisons de coins dans un ensemble monétaire.
\end{abstract}

\newpage
\tableofcontents

\newpage
\section{Introduction}

Le projet \textit{Ancient Coin Classification Algorithm for DIe Links} (ACCADIL)
a pour but de développer des outils permettant, à partir de photos de monnaies partageant les mêmes caractéristiques, de trouver facilement celles qui ont été frappées avec les mêmes coins monétaires (monnaies dites \enquote{liées}).
Il a permis le développement de deux principaux outils:

\begin{itemize}
    \item \textbf{Un algorithme de traitement d'image}
    permettant la superposition et l'identification automatique des 
    paires de monnaies qui ont le plus de chances d'être
    des liaisons de coins, c'est-à-dire d'être 
    liées par un coin monétaire \cite{labedan2025highaccuracyssimbasedscoringcoin},
    \item \textbf{Une Interface Homme-Machine (IHM) en ligne} offrant
des outils interactifs pour la vérification 
visuelle des liaisons de coins,
basée sur l'algorithme de traitement d'image.
\end{itemize}

\begin{figure}[H]
    \centering
    	\includegraphics[width=0.80\columnwidth]{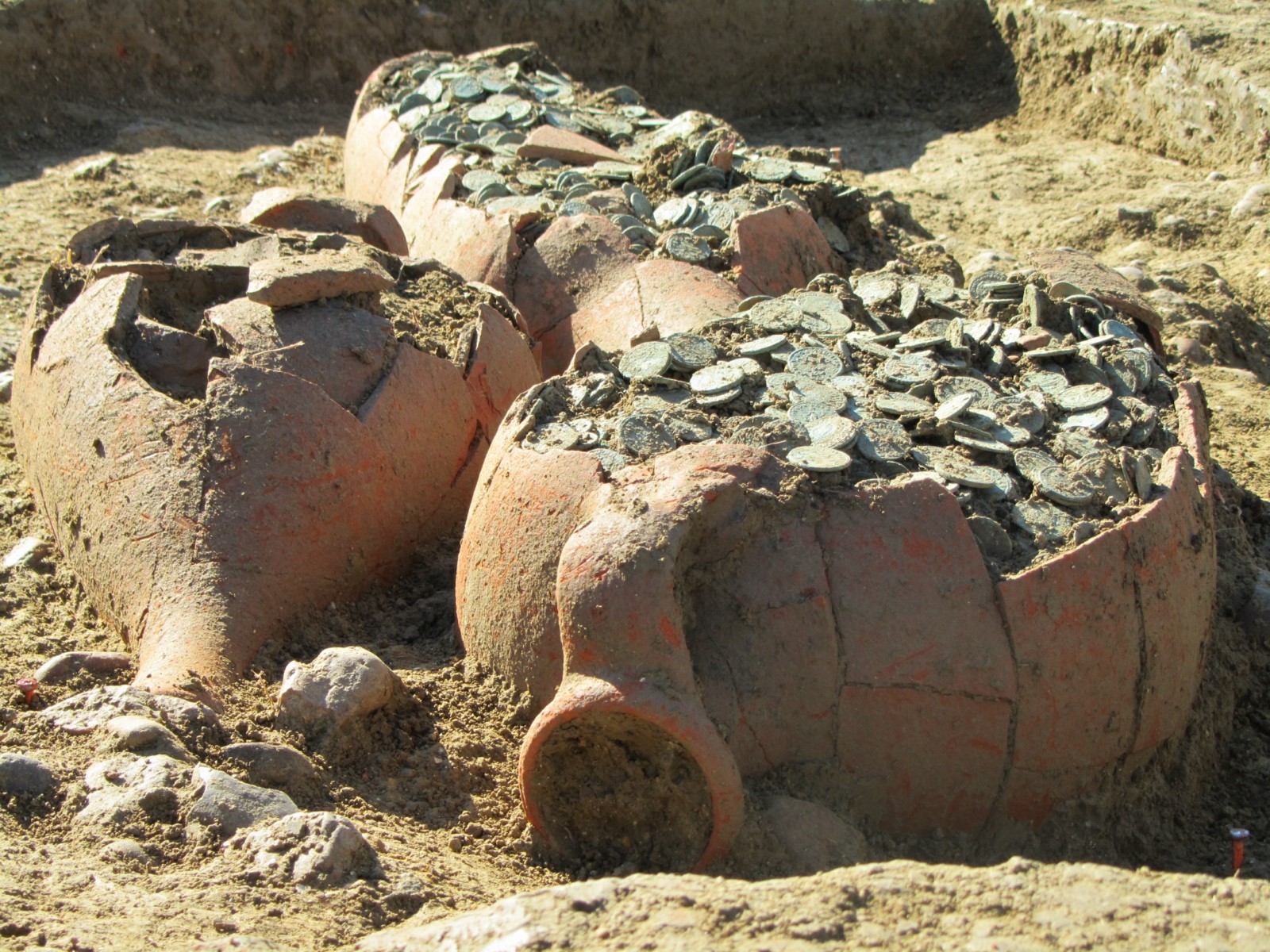}
    	\caption{Fouilles du trésor de L’Isle-Jourdain (Gers, France).}
    	\label{fig:depot_fouille}
\end{figure}

Les développements et 
les tests de ces outils
ont été effectués à partir des jeux de données \cite{AFRCBK_2024, LYCQIT_2025} issus de l'étude du dépôt monétaire de L'Isle-Jourdain 
\cite{Dieulafait2014Juillac, dieulafait2024juillacbm}
dont une photo des fouilles se trouve Figure \ref{fig:depot_fouille}. 
En  2024 et 2025, les travaux ont bénéficié du soutien d’Hadès, bureau d'investigations archéologiques. Enfin, depuis le début du projet en 2022, l'expertise numismatique a été assurée par Francis Dieulafait, numismate.

\subsection{Algorithme de traitement d'image}

\begin{figure}[H]
    \centering
    	\includegraphics[width=0.99\columnwidth]{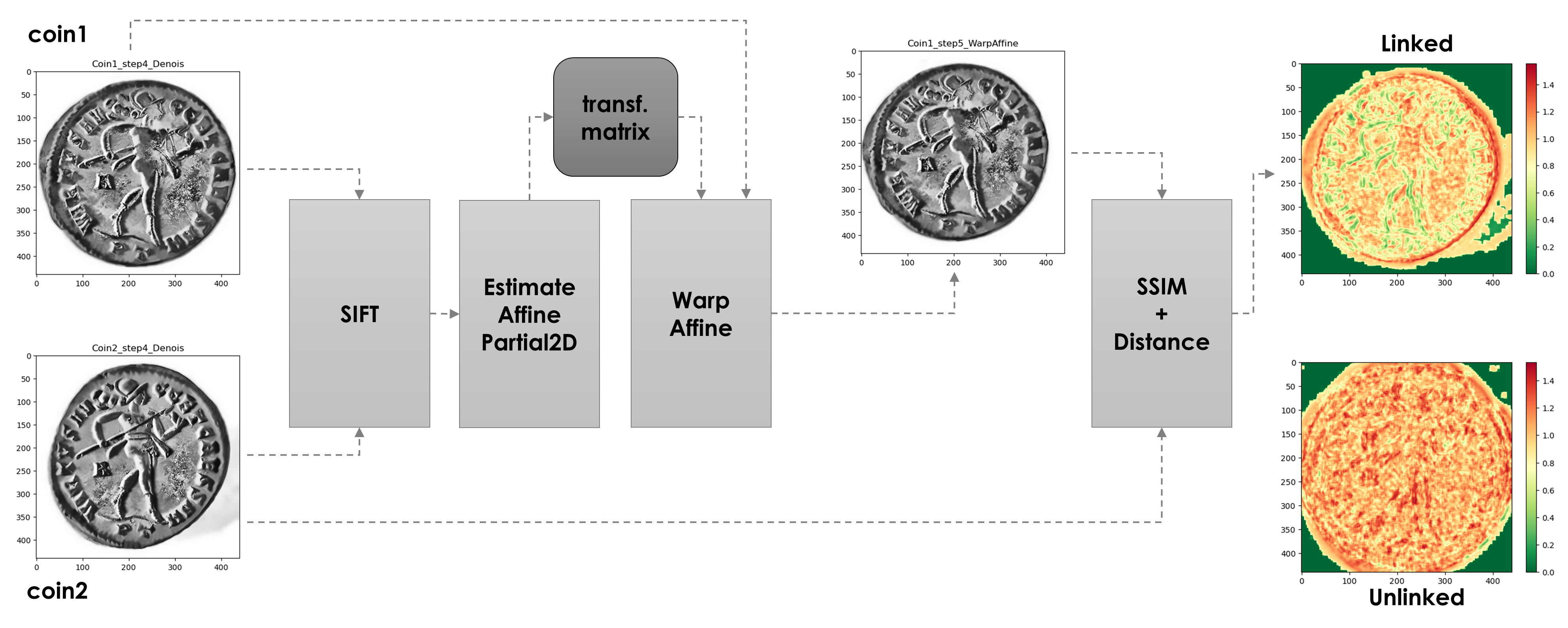}
    	\caption{Algorithme de traitement d'image du projet ACCADIL \cite{labedan2025highaccuracyssimbasedscoringcoin}.}
    	\label{fig:SSIM_GL_BL}
\end{figure}

L'algorithme de traitement d'image prend en entrée un ensemble de photos de monnaies et calcule pour chaque paire une valeur de
\textbf{distance}, comprise entre $0$ et $1$ (Figure \ref{fig:SSIM_GL_BL}). Plus cette distance est proche de $0$, plus la probabilité de liaison est importante. Inversement, plus la distance calculée est proche de $1$, plus cette probabilité est faible. La limite entre les deux ensembles (de paires liées et de paires non liées) est variable en fonction 
du contexte~:
qualité des monnaies, qualité des prises de vue, similarité des éclairages, etc.

\begin{figure}[H]
    \centering
    	\includegraphics[width=0.50\columnwidth]{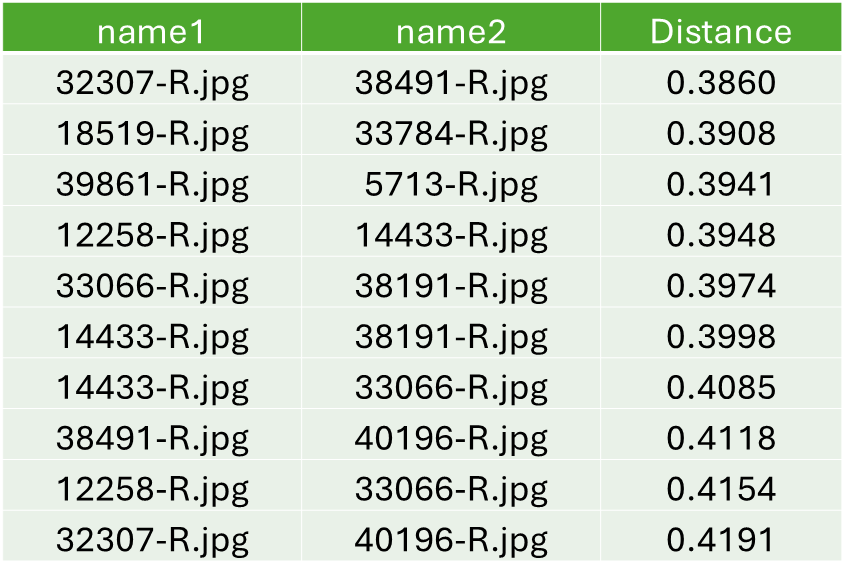}
        \caption{Classement des paires d'images suivant des distances ascendantes.}
    	\label{fig:list_results_algo2}
\end{figure}

L’algorithme classe ensuite les paires de monnaies suivant les distances ascendantes. Les paires susceptibles d’être liées sont ainsi en tête de liste (Figure \ref{fig:list_results_algo2}), directement accessibles 
pour une vérification visuelle avec l'IHM d'ACCADIL et les interactions présentées en Section \ref{resultats}.
Une courbe représentant les différentes valeurs de distances classées par ordre croissant est aussi affichée
(voir Section \ref{label_graphe_distance} et Figure \ref{fig:graphe_distances}).

\subsection{Accès à l'interface en ligne}

\subsubsection{Site internet}

Le lien du site est le suivant~:
\url{https://www.accadil.fr}\footnote{Cette adresse pourra être amenée à évoluer.}.

\begin{figure}[H]
    \centering
    	\fbox{\includegraphics[width=0.95\columnwidth]{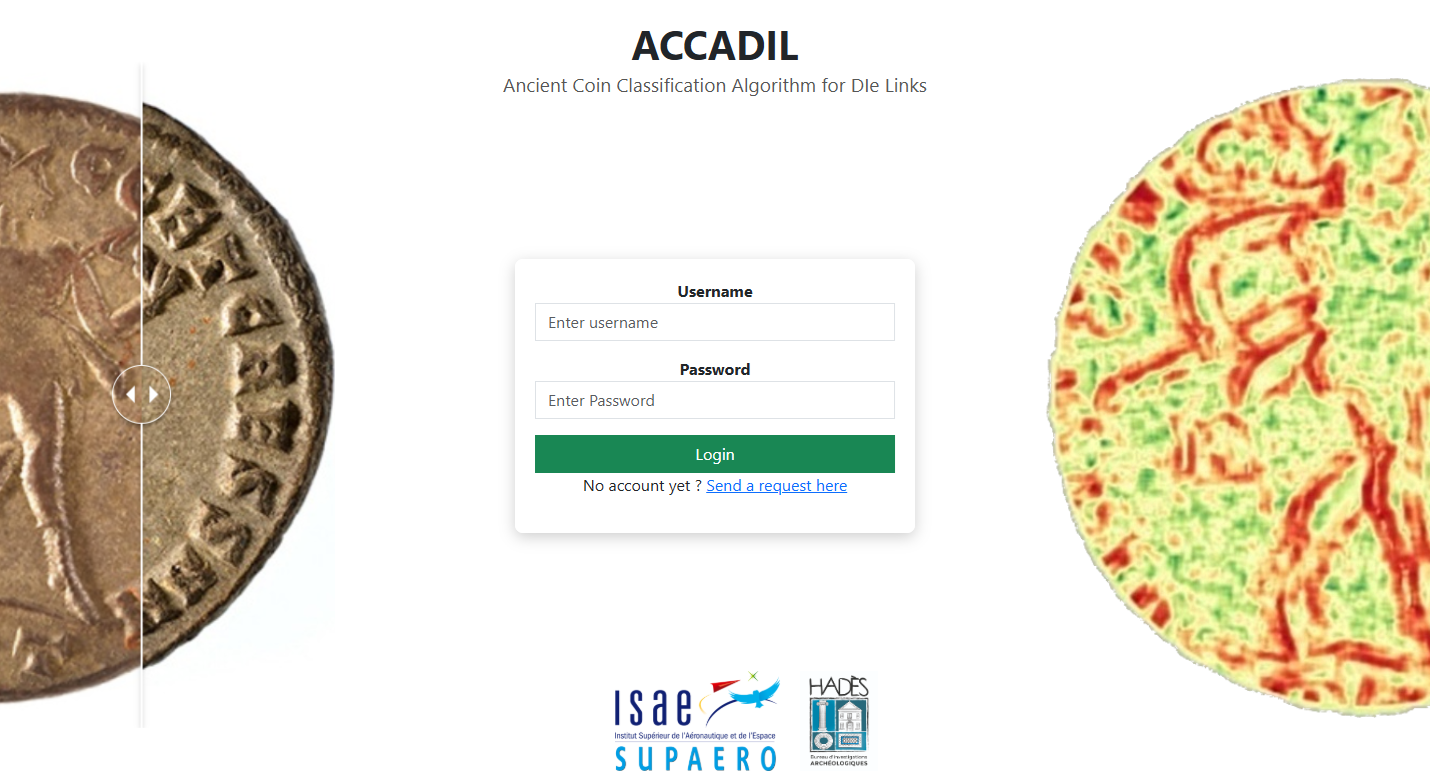}}
    	\caption{Page d'accueil 
        de l'IHM en ligne
        d'ACCADIL. }
    	\label{fig:site_accadil}
\end{figure}

\subsubsection{Inscription/Connexion}

\begin{figure}[H]
    \centering
    	\fbox{\includegraphics[width=0.45\columnwidth]{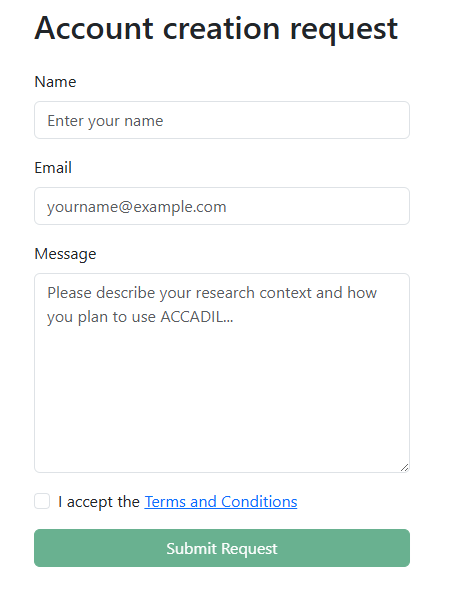}}
    	\caption{Formulaire de demande de création de compte pour utiliser l'IHM.}
    	\label{fig:account_request}
\end{figure}

Pour obtenir vos identifiants de connexion
à l'IHM en ligne,
il est nécessaire de faire une demande aux administrateurs du projet ACCADIL. Il suffit pour cela de cliquer sur
\enquote{\textit{Send a request here}}, texte de couleur bleue et souligné se trouvant sous le bouton \textit{Login} de la page d’accueil (Figure \ref{fig:site_accadil}). 
Vous pourrez ainsi remplir un formulaire (Figure \ref{fig:account_request}) avec votre nom (\textit{Name}), adresse électronique (\textit{Email}), et la description succincte de votre projet,
puis lire et accepter les (très courts) termes et conditions d’utilisation, et enfin nous envoyer votre requête en cliquant sur le bouton vert \enquote{\textit{Submit Request}}. 

En utilisant le nom d'utilisateur (\textit{Username}) et le mot de passe (\textit{Password}) transmis par les administrateurs, vous accéderez alors à votre tableau de bord (\textit{Dashboard}), présenté Section \ref{dashboard}.

\subsection{Citations}

Si vous utilisez les outils du projet ACCADIL, n'oubliez pas de citer les deux références précisées ci-dessous.

\subsubsection{Algorithme de traitement d'image}

Référence \BibTeX \ à utiliser pour l’algorithme de traitement d'image  \cite{labedan2025highaccuracyssimbasedscoringcoin} :

\begin{lstlisting}[style=accadil, language=BibTeX]
@article{labedan2025high,
  title={A High-Accuracy SSIM-based Scoring System for Coin Die Link Identification},
  author={Labedan, Patrice and Drougard, Nicolas and Berezin, Alexandre and Sun, Guowei and Dieulafait, Francis},
  journal={arXiv preprint arXiv:2502.01186},
  year={2025}
}
\end{lstlisting}

\subsubsection{Interface Homme-Machine}

Cette documentation est déposée sur arXiv. La référence bibliographique à utiliser pour la citation est directement accessible sur la page du dépôt.

\section{Le \textit{Dashboard} : description générale}
\label{dashboard}

\begin{figure}[H]
    \centering
    	\fbox{\includegraphics[width=0.90\columnwidth]{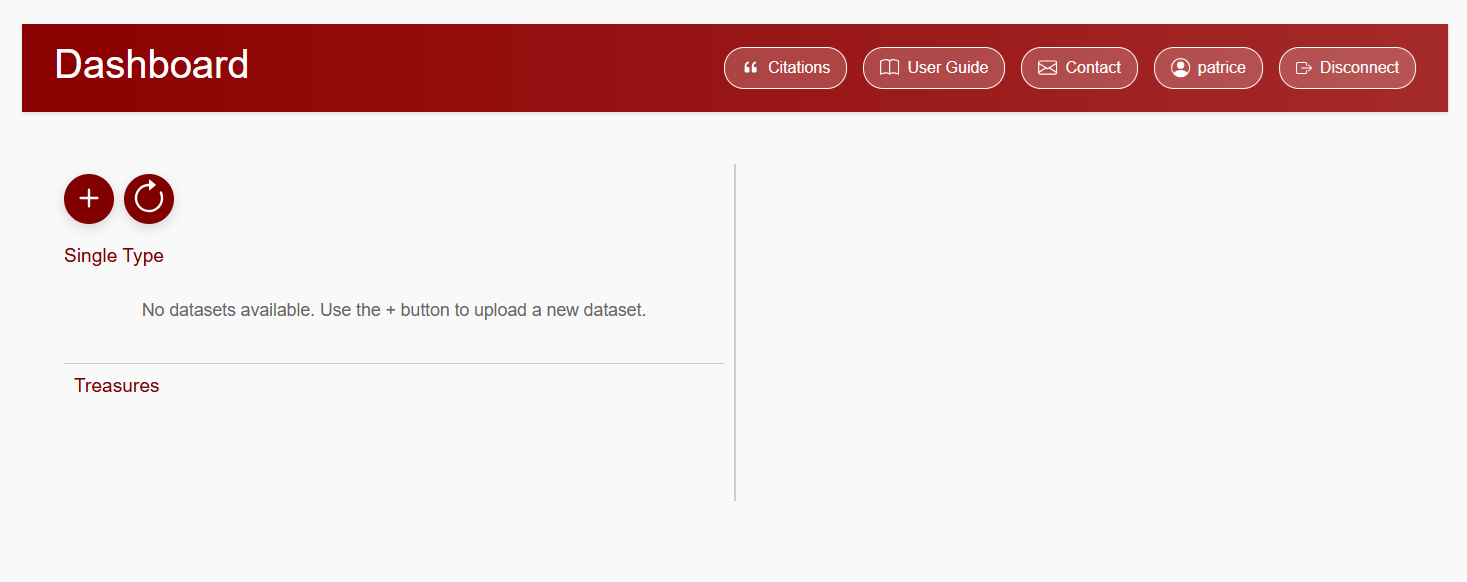}}
    	\caption{\textit{Dashboard} lors de la première connexion}
    	\label{fig:dashboard_vide}
\end{figure}

Le \enquote{\textit{Dashboard}} (tableau de bord) est votre espace de gestion des données (Figure \ref{fig:dashboard_vide}). Il contient un \textbf{bandeau} principal, au-dessus de deux zones : l'ensemble des \textbf{jeux de données} (à gauche), \textbf{les caractéristiques} du jeu de données sélectionné (à droite).

\subsection{Bandeau}

Ce bandeau vous propose plusieurs menus : 
\begin{itemize}
    \item \textbf{\textit{Citations}} : informations nécessaires pour citer les travaux d'ACCADIL si vous utilisez les résultats dans vos travaux/documents,
    \item \textbf{\textit{User Guide}} : documentation pour les utilisateurs (ce document),
    \item \textbf{\textit{Contact}} : informations pour contacter les administrateurs de l'IHM,
    \item Un rappel du nom de l'utilisateur,
    \item \textbf{\textit{Disconnect}} : bouton de déconnexion.
\end{itemize}

\subsubsection{Contact}
\label{contacts}

L'adresse de messagerie électronique pour contacter les administrateurs se trouve sur la page \textit{Dashboard}, dans le menu \textit{\enquote{Contact}}. Actuellement, il s'agit de l'adresse suivante~: \textbf{accadil.project@gmail.com}\footnote{Cette adresse pourra être amenée à changer.}.

\subsection{Zone gauche : les jeux de données}

Deux listes sont ici disponibles~: 

\begin{itemize}
    \item \textbf{\textit{Single type}}~: les ensembles d'images ne contenant qu'un seul \enquote{type} de monnaies (ensembles à \enquote{type} unique, dont le sens est défini Section \ref{prepa_donnees_def_type}).
    \item \textbf{\textit{Treasures}}~: Les ensembles d'images contenant des monnaies de types différents (ensembles à types multiples). 
\end{itemize}

Lors de votre première connexion, ces deux listes seront vides (Figure \ref{fig:dashboard_vide}).
La section \ref{ajout_donnees_global} est dédiée à l'ajout de jeux de données.

\subsection{Zone droite : les caractéristiques}

Cette zone permet d’afficher les caractéristiques du jeu de données sélectionné, comme le nombre d'images de monnaies, le nombre de paires, et l'évaluation courante des liaisons de coin. Si aucun jeu n'est sélectionné, cette zone est vide (Figure \ref{fig:dashboard_1ds_computed_selected_0}).

\section{Ajout de jeux de données}
\label{ajout_donnees_global}

L'interface permet actuellement aux utilisateurs de lancer l'algorithme de traitement d'image uniquement sur un seul type de monnaie à la fois, c'est-à-dire un ensemble de monnaies ayant les mêmes caractéristiques (Section \ref{prepa_donnees_def_type}). Le dépôt d'un type simple est expliqué Section \ref{ajout_donnees_details_un_type}.

Le dépôt de plusieurs ensembles de monnaies par l'utilisateur n'est pas possible dans la version actuelle. Toutefois, il sera possible de demander l'analyse de plusieurs types en contactant les administrateurs (Section \ref{ajout_donnees_details_plusieurs_types}).

\subsection{Définition de \textit{Type}}
\label{prepa_donnees_def_type}

\begin{definition}
Un type est un groupe de monnaies partageant la même description. Il ne concerne qu’une seule des faces de la monnaie (soit le \textbf{revers}, soit le \textbf{droit}\footnote{En numismatique, les termes \textbf{droit} et \textbf{revers} désignent les deux faces d’une monnaie (ou d’une médaille).}). Seules les monnaies du même type peuvent être potentiellement issues du même coin monétaire.
\end{definition}

\paragraph{Exemple} Un des types du trésor gallo-romain de l'Isle-Jourdain \cite{Dieulafait2014Juillac, dieulafait2024juillacbm} est composé de $17$ revers qui partagent les trois caractéristiques suivantes~: 
\begin{itemize}
    \item \textbf{Légende du revers} : \enquote{\textbf{VIRTVS AV-GG ET CAESS NN}}, 
    \item \textbf{Type de revers} : \enquote{\textbf{Mars 3b}}\footnote{Code interne à l’équipe scientifique pour la description de Mars.},
    \item \textbf{Marque de l'atelier} : \enquote{\textbf{A | - // PT}}.
\end{itemize}

\begin{figure}[H]
    \centering
    	\fbox{\includegraphics[width=0.90\columnwidth]{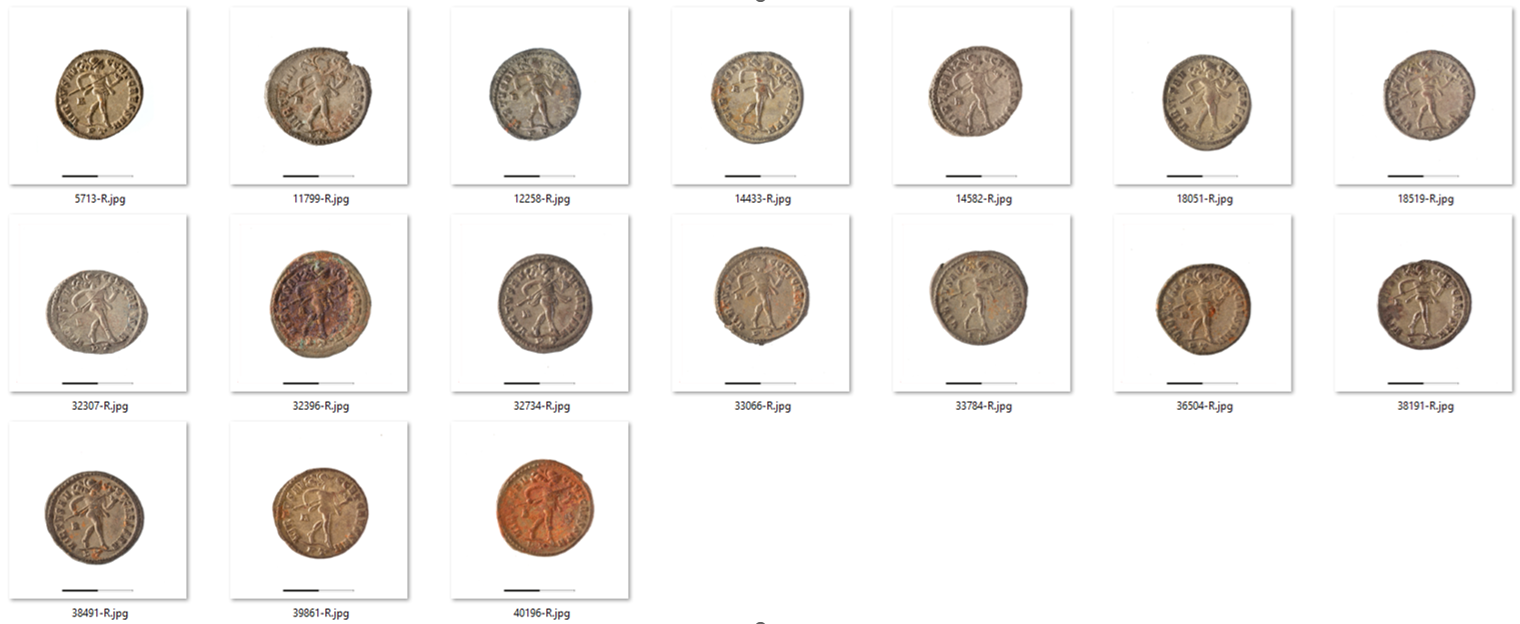}}
    	\caption{Ensemble de 17 images de revers de monnaies du même type issues du trésor de l’Isle-Jourdain.}
    	\label{fig:list_17_coins}
\end{figure}

\paragraph{Remarque}
\begin{warning}

Attention, l'interface ne trie pas les types\footnotemark. Il est nécessaire de créer un dossier par type et d'y regrouper les images du même type avant de les soumettre individuellement à l'interface (Section \ref{prepa_donnees_classement_type}).
\end{warning}

\footnotetext{Différencier automatiquement les types à partir de prises de vue représente aussi un défi pour les scientifiques. Cela ne correspond cependant pas à la problématique de l'analyse automatique des liaisons de coins traitée par le projet ACCADIL.}

\subsection{Préparation des données}
\label{prepa_donnees}

\subsubsection{Classement par \textit{Type}}
\label{prepa_donnees_classement_type}

Les algorithmes peuvent générer des temps de calculs importants, car toutes les paires possibles de monnaies sont analysées. 
Il est donc important que les images de monnaies soient \textbf{classées par type},
au sens défini dans la section précédente, pour éviter les comparaisons inutiles.
Il est aussi important de 
\textbf{n'envoyer à analyser qu'un seul type à la fois}. 
L'ensemble ne doit pas contenir un mélange de droits et de revers, mais seulement des monnaies potentiellement issues du même coin monétaire.
Le non respect de ces contraintes pourraient empêcher le bon fonctionnement de l'IHM en ligne d'ACCADIL.

\subsubsection{Taille des images}
\label{prepa_donnees_taille_images}

\begin{figure}[H]
    \centering
    	\includegraphics[width=0.55\columnwidth]{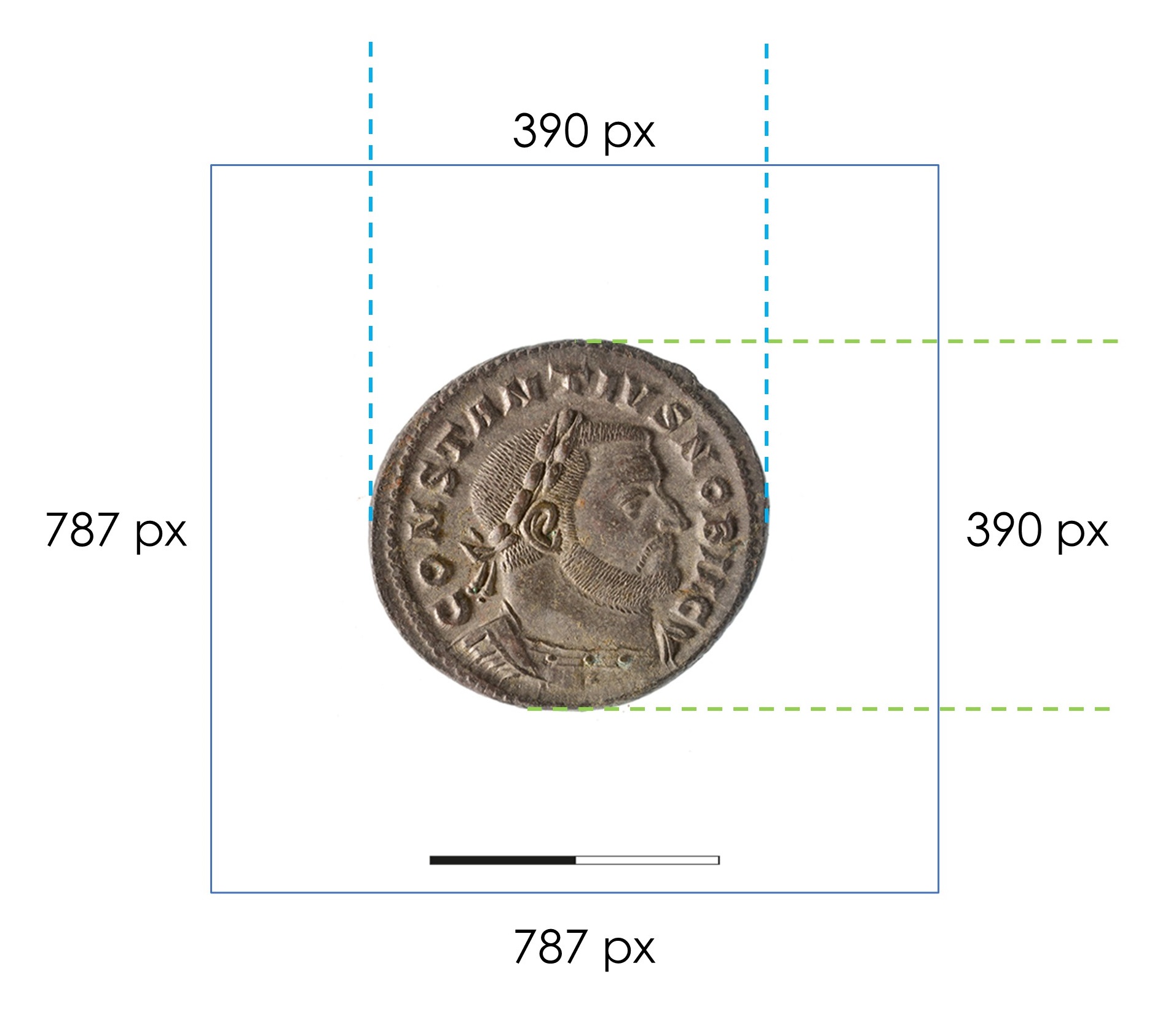}
    	\caption{
        Exemple (sur le trésor de l'Isle-Jourdain) de définition (ou taille) d'image ($787$ pixels) et de définition (ou \enquote{taille efficace}) de la monnaie ($390$ pixels).}
    	\label{fig:exemple_resolution_juillac}
\end{figure}

Pour des raisons d'optimisation des durées d'exécution, l’algorithme réduit la taille des images de telle sorte que la largeur et hauteur de la monnaie sur l'image, ou
\enquote{taille efficace}, soit environ de $400$ pixels\footnote{
Définition de la monnaie sur l'image (ou \enquote{taille efficace}, en pixels), différente de la définition de l'image, qui peut être plus importante si l'image n'est pas rognée suffisamment pour l'ajuster à la monnaie (voir Figure \ref{fig:exemple_resolution_juillac}).
} (Figure \ref{fig:exemple_resolution_juillac}). Il n'est donc pas nécessaire que les monnaies occupent un espace sur l'image (ou \enquote{taille efficace}) plus important que $400$ pixels en largeur et hauteur. 
De plus, les téléversements (\textit{Uploads}) pourraient être trop longs, voire échouer, si le jeu de données est trop volumineux. En effet, en termes d'espace de stockage, la qualité des Appareils Photos Numériques (APN) actuels mène à des images stockées sur plusieurs dizaines de mégaoctets.
Si un jeu de données contenait plusieurs centaines de telles photos, il faudrait alors téléverser plusieurs gigaoctets de données.

\subsubsection{Qualité des résultats en pratique}
\label{conseils}

\paragraph{Usure}
\label{conseils_usure}
Bien qu'une usure trop importante des monnaies peut dégrader les performances de l'algorithme,
les résultats restent satisfaisant si l’usure demeure légère.
De même, si une monnaie est cassée et qu'une partie importante est manquante,
les résultats de l'algorithme ne seront pas fiables.
Il est donc important d'interpréter les résultats en fonction de l'état des monnaies.

\paragraph{Oxydation}
\label{conseils_oxydation}

Une oxydation trop importante des monnaies peut aussi mener à des résultats peu fiables.
Ces résultats peuvent être améliorés avec un nettoyage ou une restauration des monnaies par un laboratoire spécialisé, afin d’éliminer les traces d’oxydation et de photographier des monnaies en meilleur état.

\paragraph{Éclairage}
\label{conseils_eclairage}

\begin{figure}
    \centering
    	\includegraphics[width=0.95\columnwidth]{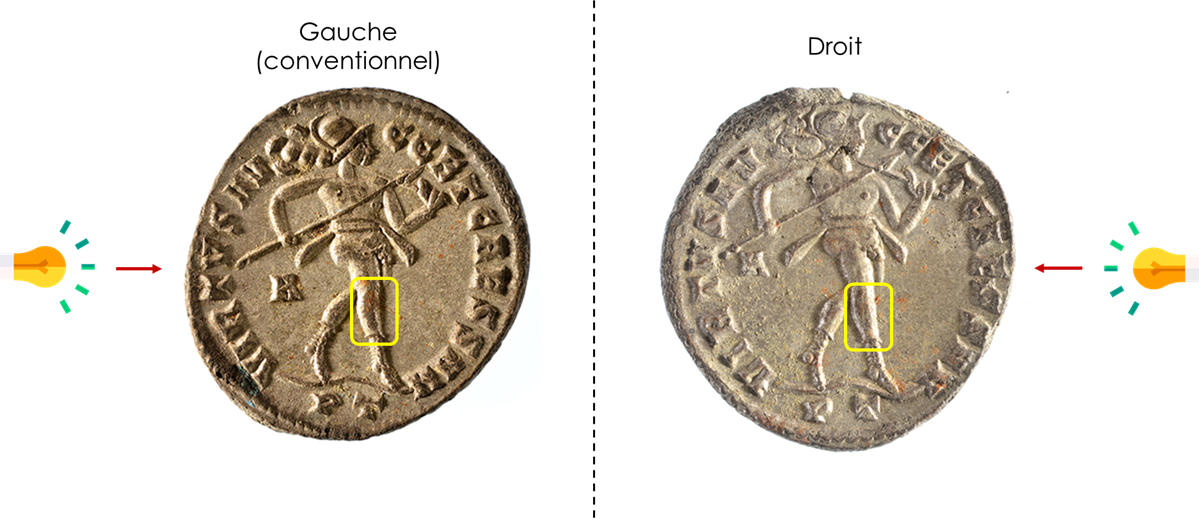}
    	\caption{Importance de l’éclairage~: 
        l'appariement des points d'intérêts de ces deux images peut être difficile pour l'algorithme car les parties claires et sombres sont inversées, comme le mollet et le tibia dans la zone encadrée en jaune.
        }
    	\label{fig:eclairage_monnaie}
\end{figure}

Ce paramètre est un élément \textbf{crucial} lors de la prise de vue des monnaies. L'algorithme étant basé sur la détection de points d’intérêt sur l'image, il est important de s'assurer que toutes les monnaies du jeu de données à analyser soient soumises au même éclairage. 
En effet, si deux photos de monnaies ont des éclairages différents, la détection des points d’intérêts mène généralement à de mauvais résultats (Figure \ref{fig:eclairage_monnaie}).

D'autre part, il est plutôt d’usage en numismatique de placer la source lumineuse à gauche. 
Ainsi, effectuer des prises de vue de monnaies avec un éclairage à gauche est intéressant pour permettre des comparaisons avec de nombreuses bases de données. 
Il existe cependant d'autres éclairages, comme l' éclairage \enquote{zenith}, visible sur certains sites de vente en ligne. 
Encore une fois, le plus important est d'avoir un éclairage uniforme sur toutes les prises de vue de monnaies dans l'ensemble à analyser.

\subsection{Ajout d'un jeu de données}
\label{ajout_donnees_details}

Sur l'interface en ligne, un ensemble de photos (de monnaies du même type) peut être envoyé à l'algorithme, en respectant certaines consignes (détaillées dans la Section \ref{ajout_donnees_details_un_type}). 
Il n'est cependant pas possible de déposer des types différents en une seule fois (comme un trésor ou un lot complet), mais il suffit de contacter les administrateurs dans l'éventualité d'un tel besoin (voir Section \ref{ajout_donnees_details_plusieurs_types}).

\subsubsection{Un seul type}
\label{ajout_donnees_details_un_type}

\begin{figure}
  \centering
  \includegraphics[width=0.35\textwidth]{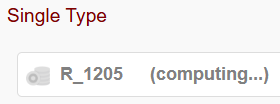}
  \hspace{3cm}
  \includegraphics[width=0.35\textwidth]{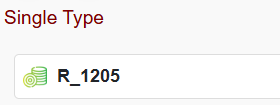}
  \caption{
  Le nom de l'ensemble de données (photos de monnaies du même type) envoyé à l'algorithme est affiché dans la zone gauche du \textit{Dashboard}~:  
  ici le nom du type est R\_1205, et il apparait comme à gauche de la figure
  lorsque le traitement des données est en cours
(grisé, accompagné de \enquote{\textit{computing}...}),
et comme à droite de la figure lorsque le traitement est terminé.}
  \label{fig:comparaison_computing_computed}
\end{figure}

Un clic sur le bouton \enquote{\textbf{+}} du \textit{Dashboard} (Figure \ref{fig:dashboard_vide}) permet de téléverser un type simple, qui devra être au \textbf{format .zip}.
Si les consignes ci-dessous ont été respectées, la prochaine connexion sur l'IHM d'ACCADIL mène à un \textit{Dashboard} qui affiche le jeu de données comme étant en cours d’évaluation (accompagné de \enquote{\textit{computing...}}), ou bien au jeu de données entièrement analysé avec les résultats disponibles (Figure \ref{fig:comparaison_computing_computed}).

\paragraph{Consignes à respecter}
\begin{warning}

\begin{itemize}
    \item \uline{Taille de l'archive .zip}~: 
    inférieure à $500$ Mo,
    \item \uline{Fichiers contenus dans l'archive}~: uniquement des images,
    \item \uline{Nombre de fichiers dans l'archive}~: inférieur à $500$,
    \item 
    Encore une fois, uniquement des monnaies du même type.
\end{itemize}

\end{warning}

\subsubsection{Plusieurs types}
\label{ajout_donnees_details_plusieurs_types}

L'analyse d'un dossier contenant plusieurs types n'est pas possible directement depuis l'interface, essentiellement car le volume d'images pourrait être trop important pour un téléversement classique. Toutefois, étudier un lot de plusieurs types (ou un \textit{trésor}) est possible~:
il suffit pour cela de contacter les administrateurs et leur détailler le projet/besoin (voir \enquote{contacts} Section \ref{contacts}).

\section{Résultats}
\label{resultats}

\subsection{Sur la page \enquote{\textit{Dashboard}}}
\label{dashboard}

\begin{figure}[H]
    \centering
    	\fbox{\includegraphics[width=0.9\columnwidth]{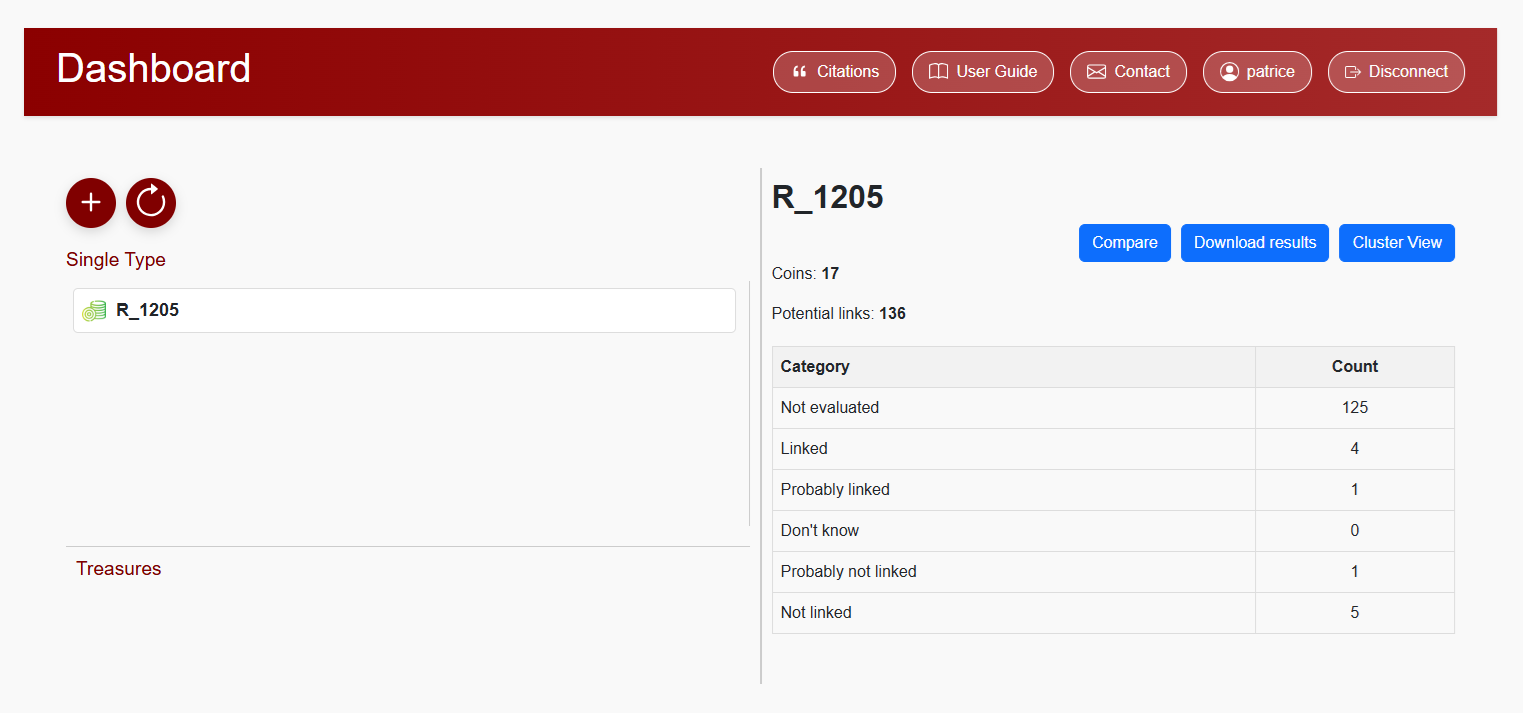}}
    	\caption{\textit{Dashboard} lorsqu'un jeu de données dont le traitement est terminé a été sélectionné (à gauche)~: ses
        caractéristiques 
        apparaissent alors à droite.}
    	\label{fig:dashboard_1ds_computed_selected_0}
\end{figure}

En cliquant sur le nom d'un type disponible\footnote{Un type \textbf{disponible} est un jeu de données pour lequel les calculs sont terminés.}, les caractéristiques de ce jeu de données
s'affichent sur la zone droite de l'écran (Figure \ref{fig:dashboard_1ds_computed_selected_0}). Les différentes informations sont :
\begin{itemize}
    \item \textit{\textbf{Coins}} : le nombre de monnaies du jeu de données ($N$),
    \item \textit{\textbf{Potential links}} : le nombre total de paires ($N_b$), avec \[N_b = \frac{N(N-1)}{2},\]
    \item \textit{\textbf{Category}} : le bilan de 
    l'analyse visuelle
    du jeu de données avec la liste des six catégories possibles, 
    c'est-à-dire les paires
    \enquote{non évaluées} (\enquote{\textit{Not evaluated}}),
    \enquote{liées} (\enquote{\textit{Linked}}), \enquote{probablement liées} (\enquote{\textit{Probably linked}}), \enquote{impossible à évaluer} (\enquote{\textit{Don't know}}), \enquote{probablement non liées} (\enquote{\textit{Probably not linked}}), ou enfin \enquote{non liées} (\enquote{\textit{Not linked}}), voir Figure \ref{fig:evaluation_options}.
\end{itemize}

Au-dessus de ces caractéristiques, trois boutons bleus apparaissent~: le bouton
\enquote{\textit{Compare}}, permettant d'afficher les résultats de l'algorithme et d'analyser visuellement les paires de monnaies (Section \ref{page_compare}); 
le bouton \enquote{\textit{Download results}} pour télécharger les résultats de l'algorithme (Section \ref{download_results}), et enfin le bouton \enquote{\textit{Cluster View}} affichant une visualisation en deux dimensions de la proximité entre les monnaies, mise en place à partir des distances calculées par l'algorithme (Section \ref{cluster_view}).

\subsubsection{Bouton \enquote{\textit{Download results}}}
\label{download_results}

\begin{figure}[H]
    \centering
    	\includegraphics[width=0.8\columnwidth]{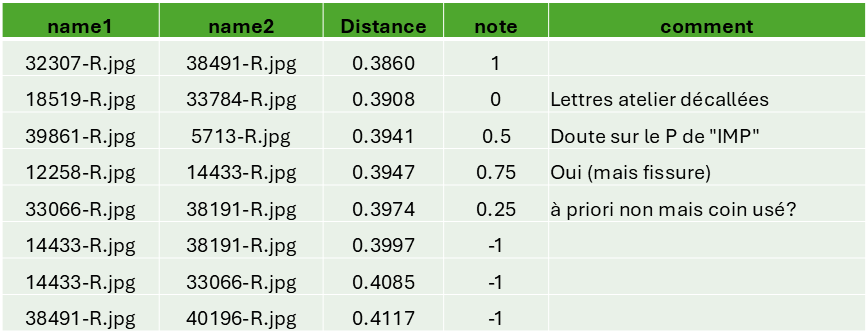}
    	\caption{Exemple de fichier
        contenant les résultats de l'analyse~: il est au format CSV,
        concerne ici
        le type \enquote{R\_1205},
        et s'appelle donc
        \enquote{\textit{notations\_R\_1205.csv}}. Pour chaque paire (correspondant aux noms de fichier présents dans les colonnes \enquote{\textit{name1}} et \enquote{\textit{name2}}), il contient la distance calculée par l'algorithme (\enquote{\textit{Distance}}), la \enquote{note} correspondant à l'analyse visuelle, et l'éventuel commentaire (\enquote{\textit{comment}}) associé.}
    	\label{fig:notations_csv}
\end{figure}

Ce bouton permet de récupérer les résultats relatifs au jeu de données, au format CSV~: 
pour chaque paire de monnaies, il contient l'évaluation (ou \enquote{\textit{note}}, parmi les catégories listées plus haut, Section \ref{dashboard} \enquote{\textit{Category}}), le commentaire (\enquote{\textit{comment}}), et la distance calculée par l'algorithme
(Figure \ref{fig:notations_csv}).

\subsubsection{Bouton \enquote{\textit{Cluster view}}}
\label{cluster_view}

\begin{figure}[H]
    \centering
    	\fbox{\includegraphics[width=0.9\columnwidth]{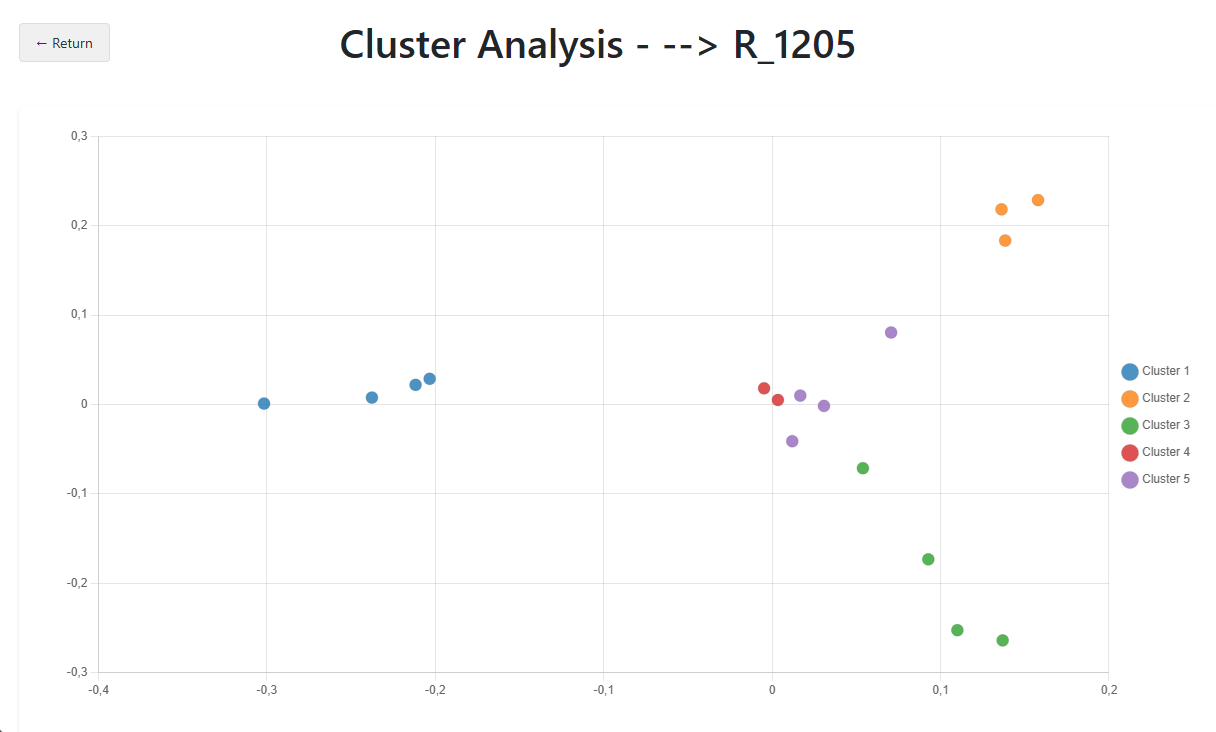}}
    	\caption{Visualisation en deux dimensions des distances entre les photos de monnaies, avec les résultats temporaires des algorithmes de \textit{clustering}, disponibles sur la page \enquote{\textit{Cluster view}}.}
    	\label{fig:clusters_ds8}
\end{figure}

Cette page permet de visualiser en deux dimensions les distances calculées par l'algorithme, ainsi que les résultats des algorithmes de \textit{clustering}. 
Ces résultats ne sont pas présents dans le fichier qui peut être téléchargé avec le bouton \textit{Download results}.
Sur l'exemple de la figure \ref{fig:clusters_ds8} (jeu de données \textit{R\_1205}), on remarque que les points de la même couleur pourraient être des monnaies liées par un coin (car ils sont proches dans cette visualisation).

Les outils permettant de générer les résultats de cette page sont en cours de développement. Ces derniers ne doivent donc pas encore être pris en compte pour le moment. 

\subsection{Sur la page \enquote{\textit{Compare}}}
\label{page_compare}

Il s'agit de la page principale pour l'affichage et l'évaluation des résultats obtenus par l'algorithme de traitement d'image. Différentes interactions sont proposées pour la vérification visuelle des liaisons.  
Les sections suivantes se concentrent sur la description de cette page et des interactions proposées.

\begin{figure}[H]
    \centering
    	\fbox{\includegraphics[width=0.99\columnwidth]{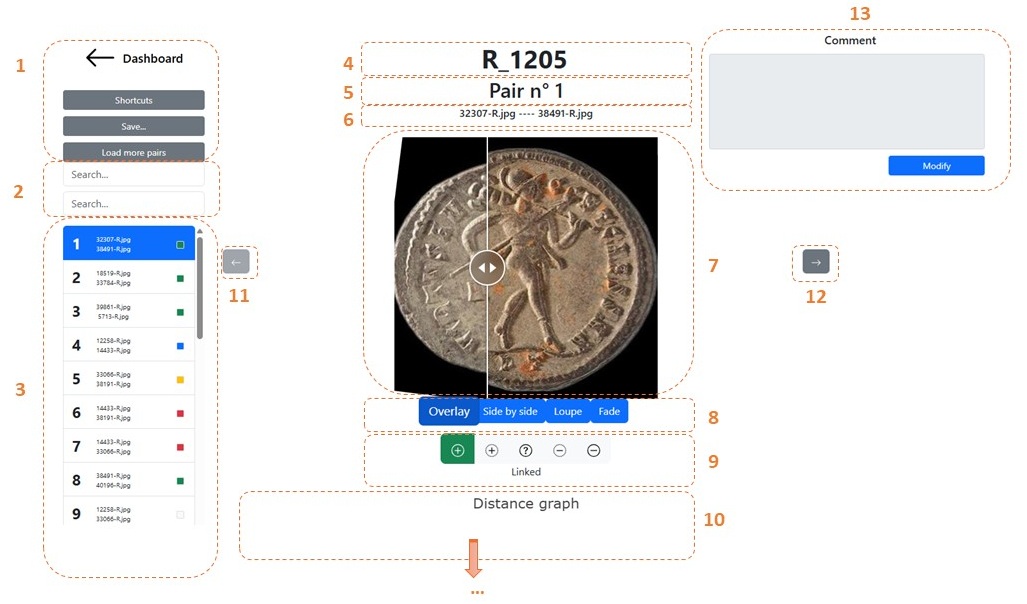}}
    	\caption{Différentes zones de la page \textit{Compare} 
        numérotées et présentées dans le tableau \ref{tab:zones_compare}
        (jeu de données \enquote{R\_1205}). 
        }
    	\label{fig:zones_compare}
\end{figure}

        \begin{table}[H]
            \centering
            \caption{ 
            Signification des différentes zones de la page \enquote{\textit{Compare}}.}
            \label{tab:zones_compare}
            \begin{tabular}{ll}
                \toprule
                \textbf{Zone} & \textbf{Signification} \\
                \midrule
                1 & Retour au tableau de bord (\enquote{$\leftarrow$ \textbf{\textit{Dashboard}}})~; \\
                  & Rappel des raccourcis clavier (\enquote{\textbf{\textit{Shortcuts}}})~; \\
                  & Enregistrement 
                  du travail
                  en cours
                  (\enquote{\textbf{\textit{Save}...}})~; \\
                  & Chargement des $50$ paires suivantes (\enquote{\textbf{\textit{Load more pairs}}}). \\
                2 & Recherche de paires à partir de noms de fichiers. \\
                3 & Liste des paires classées
                des liaisons les
                plus probables 
                aux moins probables\\ 
                & (avec 
                  les noms des deux fichiers, et l'évaluation, ou \enquote{note}, associée). \\
                4 & Nom du jeu de données, ou \enquote{type}. \\
                5 & Numéro de la paire de monnaies visualisée. \\
                6 & Le nom des deux fichiers  
                de la paire courante. \\
                7 & Les images des deux monnaies de la paire courante. \\
                8 & Choix du mode~: \enquote{\textit{Overlay}}, \enquote{\textit{Side by side}}, \enquote{\textit{Loupe}} ou \enquote{\textit{Fade} }\\ 
                & (voir Section \ref{modevisu}). \\
                9 & Evaluation de l’utilisateur, ou \enquote{note} parmi cinq disponibles~: paire liée,\\
                  &  
                  probablement liée, impossible à évaluer, probablement non liée, et non liée.\\
                10 & 
                Courbe des distances classées dans l'ordre croissant (\enquote{\textit{Distance graph}}). \\
                11 & Bouton pour l'affichage de la paire précédente. \\
                12 & Bouton pour l'affichage de la paire suivante. \\
                13 & Commentaire concernant la paire courante. \\
                \bottomrule   
            \end{tabular}
					
        \end{table}

\subsubsection{Zone 3 : Liste des paires classées}

L'algorithme classe les paires suivant une valeur ascendante de distance calculée par l'algorithme de traitement d'image. Les paires ayant les valeurs de distance les plus faibles, en tête de liste, ont le plus de chances d'être des liaisons (Figure \ref{fig:list_results_algo2}). 
Pour des raisons d'optimisation du temps de chargement de la page, seules les cinquante premières paires sont affichées.
Pour ajouter les cinquante paires suivantes, il suffit de cliquer sur le bouton \enquote{\textit{Load more pairs}}.

\subsubsection{Zone 9~: 
Attribution d'une \enquote{note} (évaluation)}

\begin{figure}[H]
    \centering
    	\includegraphics[width=0.4\columnwidth]{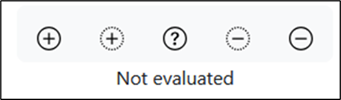}
    	\caption{Boutons permettant d'évaluer la paire de monnaies courante, c'est-à-dire de lui attribuer une \enquote{note} (zone 9 sur la figure \ref{fig:zones_compare}).
        }
    	\label{fig:evaluations}
\end{figure}

\begin{figure}[H]
    \centering
    	\includegraphics[width=0.99\columnwidth]{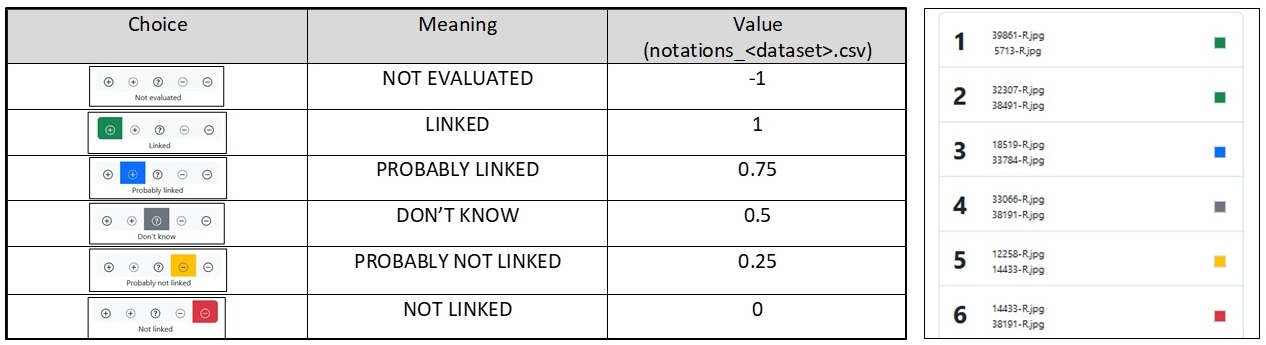}
    	\caption{Evaluations~: à gauche les différentes options (ou \enquote{catégories}) pour noter les paires, avec les valeurs associées (\textit{value}), et 
        à droite,  
        les notes sélectionnées par l'utilisateur 
        rappelées par des couleurs
        dans la liste des paires
        (zone 3 dans la figure \ref{fig:zones_compare}).}
    	\label{fig:evaluation_options}
\end{figure}

Par défaut, chaque paire de monnaies est considérée comme \enquote{\textit{Not Evaluated}}. Ensuite, il est possible de changer l'évaluation, c'est-à-dire la \enquote{note}, en utilisant les boutons sous la zone d'affichage des images et les boutons bleus concernant le mode de visualisation (Figure \ref{fig:evaluations} et \ref{fig:evaluation_options}). 

\subsubsection{Zone 8 : Choix du mode de visualisation}
\label{modevisu}

Quatre modes de visualisation sont disponibles : \enquote{\textit{Overlay}}, \enquote{\textit{Side by side}}, \enquote{\textit{Loupe}}, et \enquote{\textit{Fade}}.

\paragraph{Mode \enquote{\textit{Overlay}}}

\begin{figure}[H]
    \centering
    	\fbox{\includegraphics[width=0.5\columnwidth]{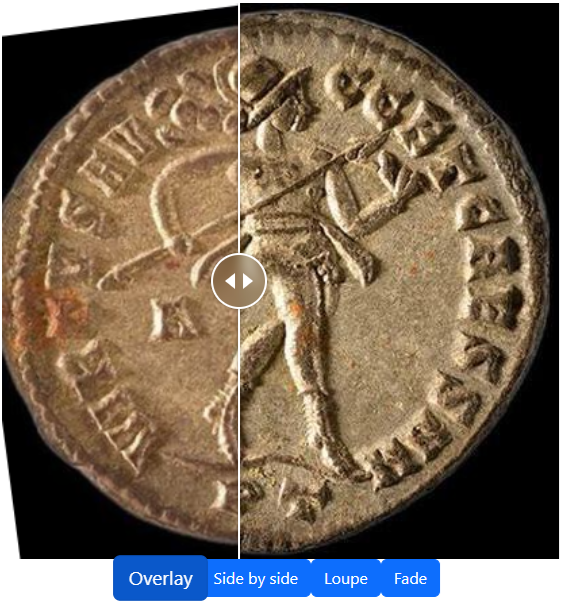}}
    	\caption{Mode \enquote{\textit{Overlay}}.}
    	\label{fig:compare_mode1_overlay}
\end{figure}

Une barre verticale sépare les deux monnaies qui se superposent. Elle peut être déplacée vers la gauche ou vers la droite pour vérifier si les monnaies \enquote{jointent} (Figure \ref{fig:compare_mode1_overlay}).

\paragraph{Mode \enquote{\textit{Side by side}}}

\begin{figure}[H]
    \centering
    	\fbox{\includegraphics[width=0.95\columnwidth]{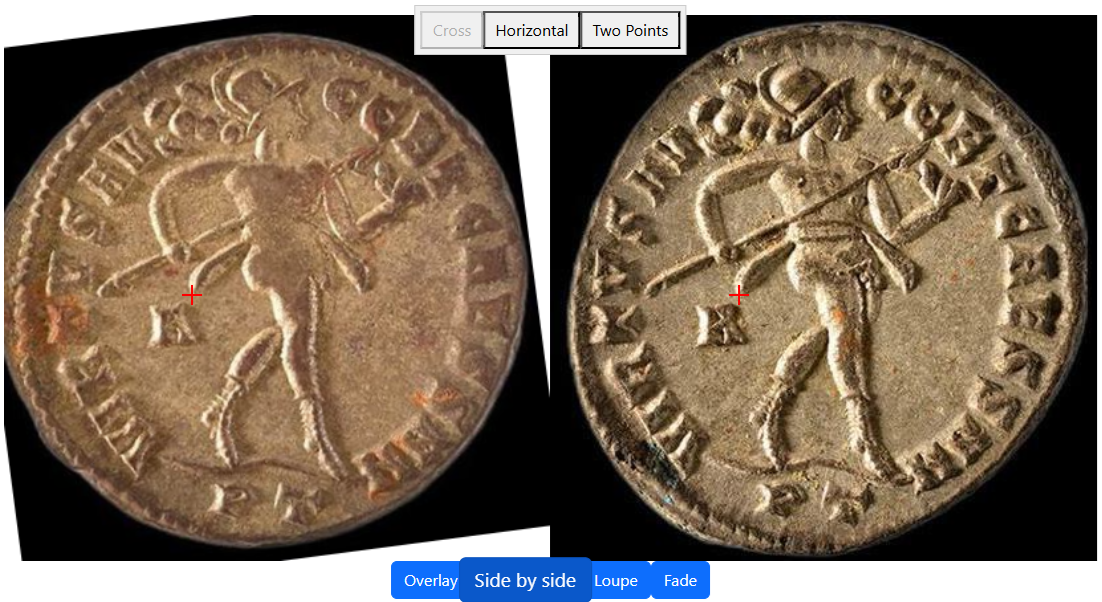}}
    	\caption{Mode \enquote{\textit{Side by side}} 
        avec le
        sous-mode \enquote{\textit{Cross}}.}
    	\label{fig:compare_mode2_side_cross}
\end{figure}

\begin{figure}[H]
    \centering
    	\fbox{\includegraphics[width=0.95\columnwidth]{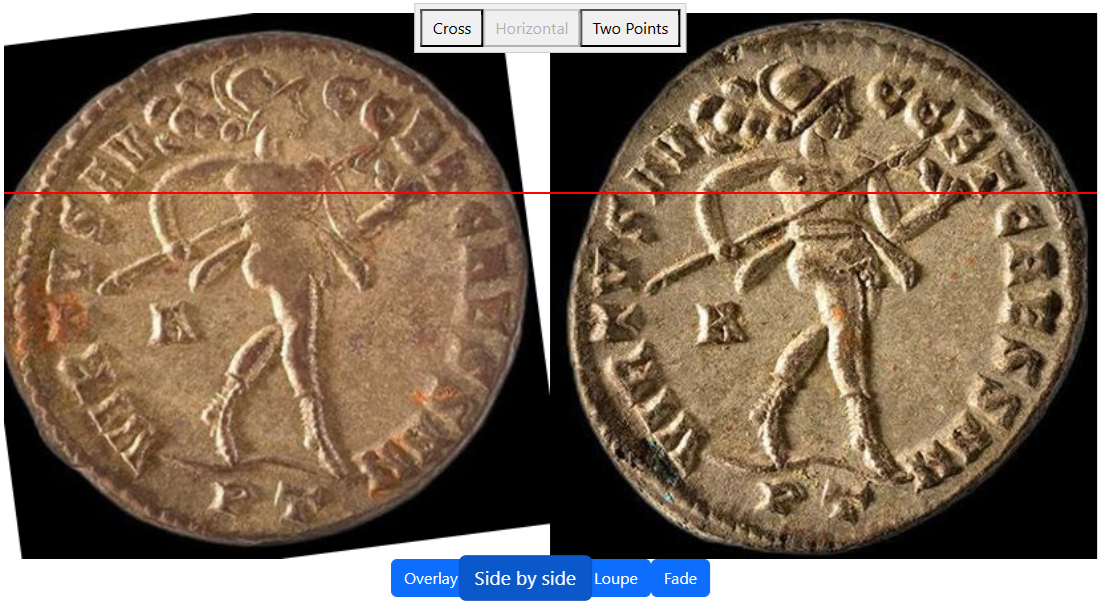}}
    	\caption{Mode \enquote{\textit{Side by side}}  
        avec le
        sous-mode \enquote{\textit{Horizontal}}.}
    	\label{fig:compare_mode2_side_barre}
\end{figure}

\begin{figure}[H]
    \centering
    	\fbox{\includegraphics[width=0.95\columnwidth]{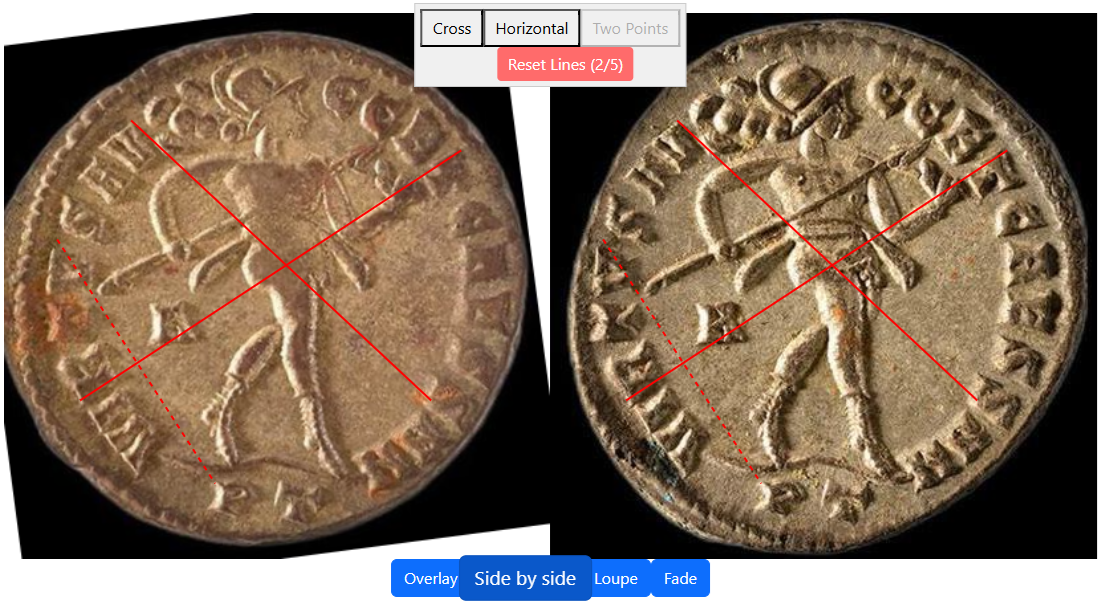}}
    	\caption{Mode \enquote{\textit{Side by side}}  
        avec le
        sous-mode \enquote{\textit{Two points}}.}
    	\label{fig:compare_mode2_side_lines}
\end{figure}

Les deux monnaies sont visibles, côte à côte, et des repères visuels sont ajoutés simultanément pour faciliter la comparaison (croix ou traits). Il existe trois sous-modes pour cet affichage. Par défaut, le sous-mode \enquote{\textit{Cross}} affiche une croix rouge simultanément sur les deux monnaies là où se trouve le curseur (Figure \ref{fig:compare_mode2_side_cross}). 
De la même manière,
le sous-mode \enquote{\textit{Horizontal}} permet 
d'afficher une barre horizontale simultanément sur les deux monnaies (Figure \ref{fig:compare_mode2_side_barre}). Enfin, le sous-mode \enquote{\textit{Two points}} permet de tracer un à cinq segments sur les deux monnaies (Figure \ref{fig:compare_mode2_side_lines}).

\paragraph{Mode \enquote{\textit{Loupe}}}

\begin{figure}[H]
    \centering
    	\fbox{\includegraphics[width=0.5\columnwidth]{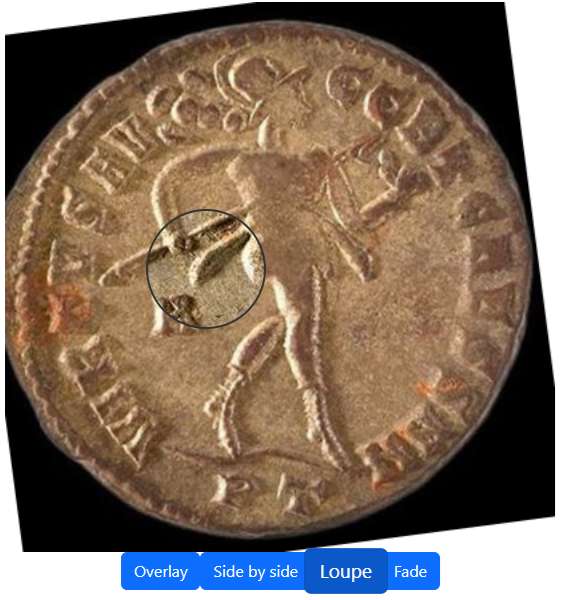}}
    	\caption{Mode \enquote{\textit{Loupe}}.}
    	\label{fig:compare_mode3_loupe}
\end{figure}

Sur l'image de la première monnaie,
la deuxième monnaie 
est affichée seulement
sur un disque (\enquote{\textit{loupe}}) centré sur le curseur. Les touches clavier \enquote{\textbf{+}} et \enquote{\textbf{-}} permettent de changer la taille de la loupe (Figure \ref{fig:compare_mode3_loupe}).

\paragraph{Mode \enquote{\textit{Fade}}}

\begin{figure}[H]
    \centering
    	\fbox{\includegraphics[width=0.65\columnwidth]{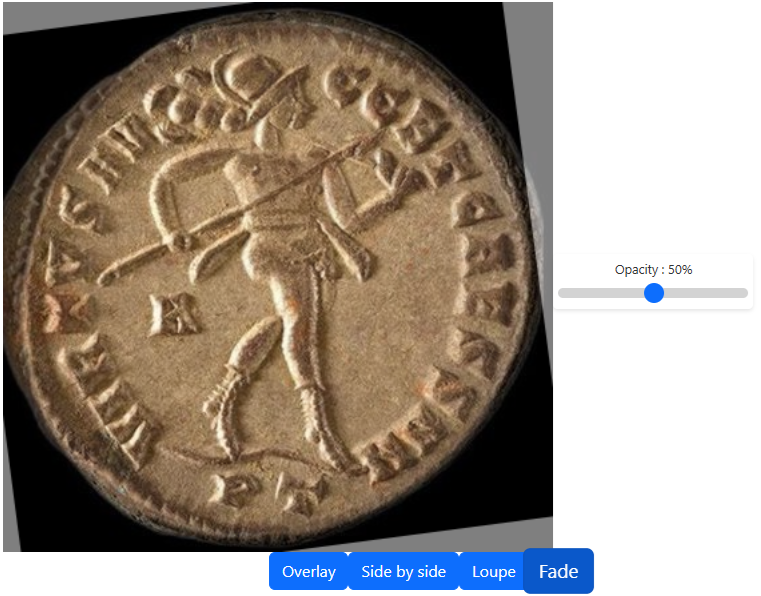}}
    	\caption{Mode \enquote{\textit{Fade}}.}
    	\label{fig:compare_mode4_fade}
\end{figure}

Ici, les monnaies se superposent, et il est possible de modifier la transparence pour visualiser plus ou moins une des monnaies.  
Il peut être utile de passer rapidement de l'une à l'autre pour se rendre compte de la bonne superposition (Figure \ref{fig:compare_mode4_fade}). 

\subsubsection{Zone 10~: Courbe des distances}
\label{label_graphe_distance}

\begin{figure}[H]
    \centering
    	\fbox{\includegraphics[width=0.90\columnwidth]{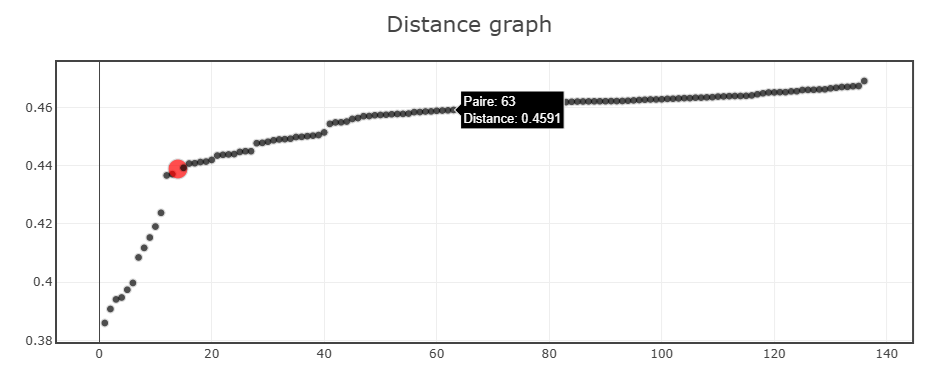}}
    	\caption{
        Courbe
        des distances classées par ordre croissant (axe x~: numéro de la paire~; axe y~: valeur de distance).  
        Le point rouge représente la paire en cours d'évaluation. 
        A gauche la courbe a une pente plus forte qu'à droite}
    	\label{fig:graphe_distances}
\end{figure}

Au bas de la page \textit{Compare}, les valeurs de distances classées dans l'ordre croissant sont affichées (Figure \ref{fig:graphe_distances}). La paire en cours d'évaluation est rappelée par le point rouge.
On peut distinguer deux zones : la première, à gauche, représente en général une courbe ascendante à forte pente, qui correspond aux paires les plus susceptibles d'être des liaisons (valeurs de distance les plus faibles). Mais à partir d'un certain seuil de distance, on constate une sorte de \enquote{cassure} dans la pente, qui devient beaucoup plus douce. Il est intéressant d'analyser visuellement les paires au moins jusqu'à cette zone de transition entre les deux pentes. 

\subsubsection{Zone 6 : Téléchargement d'une image de contrôle}

Il existe la possibilité de télécharger une visualisation courante des images, appelée \enquote{image de contrôle}, par exemple pour illustrer la note choisie pour une paire de monnaies. Cette fonctionnalité est disponible pour les modes \enquote{\textit{Side by side}}, \enquote{\textit{Loupe}}, et \enquote{\textit{Fade}} (indisponible pour le mode \enquote{\textit{Overlay}}).

\begin{figure}[H]
    \centering
    	\fbox{\includegraphics[width=0.40\columnwidth]{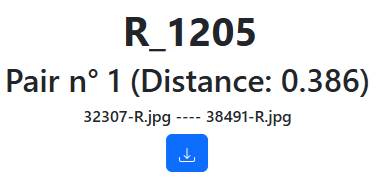}}
    	\caption{Bouton pour téléchargement d'une image de contrôle (sous la zone 6, voir Figure \ref{fig:zones_compare}).}
    	\label{fig:print_bouton}
\end{figure}

Un bouton spécifique peut être utilisé pour le mode \enquote{\textit{Fade}} (Figure \ref{fig:print_bouton}) ainsi que le sous-mode \enquote{\textit{two points}} du mode \enquote{\textit{Side by side}}. Par contre, pour tous les autres modes l'utilisation de ce bouton est problématique car le curseur se trouve à l'extérieur de la zone d'affichage des monnaies au moment de cliquer sur le bouton. 
Pour palier ce problème, le raccourci \enquote{F2} permet de télécharger l'image de contrôle 
avec ces symboles, sans avoir à cliquer sur le bouton.

\begin{figure}[H]
    \centering
    	\includegraphics[width=0.99\columnwidth]{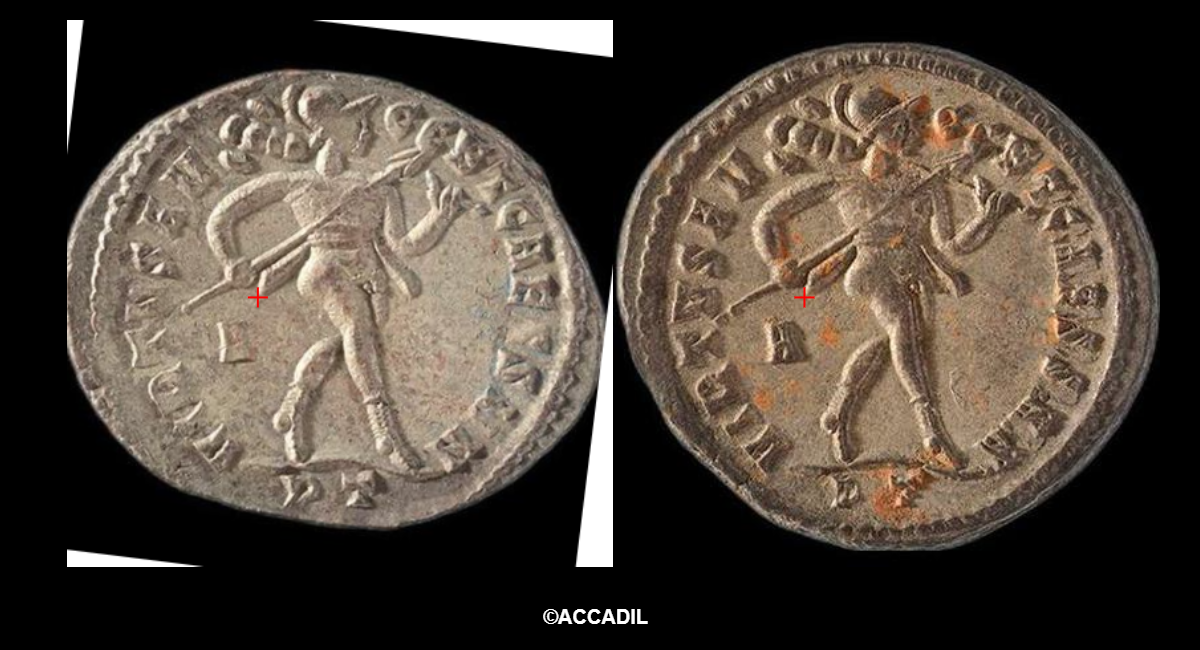}
    	\caption{Image de \enquote{contrôle} 
        d'une paire en cours d'analyse,
        avec le mode \enquote{\textit{Side by side}},
        le sous-mode \enquote{\textit{cross}}, et
        une
        croix sur les deux monnaies. Le
        nom 
        de ce 
        fichier 
        téléchargé à partir de l'interface
        est \enquote{\textit{32307-R.jpg-38491-R.jpg.png}}.
        }
    	\label{fig:print_cross}
\end{figure}

Le fichier téléchargé se trouve ensuite dans le dossier des téléchargements. Le nom de l'image 
contient le nom des deux monnaies concernées. Par exemple, pour les deux images \textit{32307-R.jpg} et \textit{38491-R.jpg}, le nom du fichier est \textit{32307-R.jpg-38491-R.jpg.png} (Figure \ref{fig:print_cross}). \\

\paragraph{Remarque}
\begin{warning}
Les images téléchargées peuvent être réutilisées à condition que le logo \enquote{(c) ACCADIL} soit bien visible (comme sur la figure \ref{fig:print_cross}). 
Pour avoir une image de qualité suffisante, le niveau de zoom de votre navigateur doit être de 100\%\footnotemark. 
S'il n'est pas à 100\%, un message l'indiquera, et l'image sera tout de même  téléchargée mais ne sera certainement pas exploitable.
\end{warning}
\footnotetext{Pour modifier le niveau de zoom~:  
touche CTRL + molette souris.}

\subsubsection{Zone 13 : Ajout de commentaires}

Il est possible d'ajouter un commentaire pour chaque paire évaluée (zone 13, Figure  \ref{fig:zones_compare}). Cela permet de préciser les raisons de la note choisie. Ces commentaires sont enregistrés dans le fichier des résultats de l'analyse qui peut être téléchargé (Section \ref{download_results}).

\begin{thebibliography}{9}

\bibitem{labedan2025highaccuracyssimbasedscoringcoin}
Patrice Labedan, Nicolas Drougard, Alexandre Berezin, Guowei Sun, and Francis Dieulafait.
\newblock {A High-Accuracy SSIM-based Scoring System for Coin Die Link Identification}, 2025.
\newblock \href{https://arxiv.org/abs/2502.01186}{arXiv:2502.01186}.

\bibitem{Dieulafait2014Juillac}
Francis Dieulafait.
\newblock {Trésor monétaire de Juillac (L’Isle‑Jourdain — 32) : 1\textsuperscript{er} rapport intermédiaire 2014}.
\newblock Projet Collectif de Recherche 2014–2016, Service régional de l’Archéologie, Région Midi‑Pyrénées, France, 2014.
\newblock Note: Avec la collaboration de S. Bach, M.-L. Berdeaux-Le Brazidec, Chr. Dieulafait, J.-M. Doyen, Fr. Fantuzzo, O. Gaiffe, V. Geneviève, A. Le Guen, B. Marty, J.-Fr. Peiré, M. Vaginay ; laboratoire \emph{Materia Viva}. O.A. n°8562; E.A.n°321600145.

\bibitem{ghey2024tetrarchic}
Eleanor Ghey.
\newblock {Recent Discoveries of Tetrarchic Hoards from Roman Britain and their Wider Context}, 2024.
\newblock British Museum, London.
\newblock \href{https://www.britishmuseumshoponline.org/rp-236-recent-discoveries-of-tetrarchic-hoards-from-roman-britain-and-their-wider-context.html}{URL}.
\newblock DOI: \href{https://doi.org/10.48582/ansj-ty90}{10.48582/ansj-ty90}.

\bibitem{dieulafait2024juillacbm}
Francis Dieulafait and Vincent Geneviève.
\newblock {The Juillac Hoard (L’Isle-Jourdain, Gers): from discovery to study}.
\newblock In: Recent Discoveries of Tetrarchic Hoards from Roman Britain and their Wider Context, edited by Eleanor Ghey et al., pp. 136--148. British Museum, London, 2024.
\newblock \href{https://www.britishmuseumshoponline.org/rp-236-recent-discoveries-of-tetrarchic-hoards-from-roman-britain-and-their-wider-context.html}{URL}.
\newblock DOI: \href{https://doi.org/10.48582/ansj-ty90}{10.48582/ansj-ty90}.

\bibitem{AFRCBK_2024}
Patrice Labedan, Nicolas Drougard, and Francis Dieulafait.
\newblock {Datasets for Accadil (V1)}, 2024.
\newblock ISAE-SUPAERO Dataverse.
\newblock DOI: \href{https://doi.org/10.34849/AFRCBK}{10.34849/AFRCBK}.

\bibitem{LYCQIT_2025}
Patrice Labedan, Nicolas Drougard, and Francis Dieulafait.
\newblock {Accadil Project: 25 Datasets for Die Link Studies}, 2025.
\newblock Recherche Data Gouv.
\newblock Version 1.0.
\newblock DOI: \href{https://doi.org/10.57745/LYCQIT}{10.57745/LYCQIT}.

\end{thebibliography}


\subsubsection*{Remerciements}

Le développement de la plateforme ACCADIL a bénéficié du soutien du bureau d’investigations archéologiques Hadès (2024–2025) et de la contribution scientifique de Francis Dieulafait (numismate) depuis 2022.

Les auteurs remercient chaleureusement les stagiaires ayant participé entre 2022 et 2025 : Alexandre Berezin, Guowei Sun, Daniel Zayat et Ralph Maatouk (ISAE-SUPAERO, cursus Ingénieur, 2ème année), Khaoula Slimani (ENAC, Master Interactions Homme-Machine), et Lilian Monestier (IUT Informatique, Blagnac).

Nos remerciements vont également aux numismates pour leur participation durant le processus de conception de l'interface : 
\begin{itemize}
    \item \textbf{Francis Dieulafait} : Rattaché UMR 5608 - équipe RHAdAMANTE \footnote{Recherches en Histoire et Archéologie des Âges des Métaux et de l'Antiquité en Europe (RHAdAMANTE)}, Laboratoire TRACES \footnote{Travaux et Recherches Archéologiques sur les Cultures, les Espaces et les Sociétés (TRACES)} - Université Toulouse - Jean Jaurès.
    \item \textbf{Jean-Marc Doyen} : Chercheur associé HDR, Unité de recherche HALMA \footnote{Histoire Archéologie Littérature des Mondes Anciens (HALMA)}, UMR 8164 (Univ. Lille, CNRS, MC).
    \item \textbf{Marie-Laure Le Brazidec} : Chercheur en Numismatique antique, Membre associé équipe RHAdAMANTE, Laboratoire TRACES - UMR 5608, Université Toulouse - Jean Jaurès.
    \item \textbf{Eneko Hiriart} : Chercheur CNRS \footnote{Centre National de la Recherche Scientifique (CNRS)}, Archeosciences Bordeaux, UMR 6034, Université Bordeaux Montaigne.
    \item \textbf{Guillaume Malingue} : Numismate, spécialiste du monnayage de Carthage et de l’Antiquité tardive.
    \item \textbf{Guillaume de Méritens de Villeneuve} : Chargé de recherches du F.R.S.-FNRS, Université de Namur, Centre Fontes Antiquitatis (Institut PaTHs).
    \item \textbf{Marianne Schreinemachers} : Numismate, Bureau d’investigations archéologiques Hadès, Thèse de doctorat portant sur la circulation monétaire entre Aquitaine et Narbonnaise à l’époque impériale romaine à partir de trois corpus gersois (2022-2025)\footnote{\href{https://theses.fr/s343998}{https://theses.fr/s343998}}.
\end{itemize}

\subsubsection*{Clichés}

\begin{itemize}
    \item Figure 1 : Patrice Labedan.
    \item Monnaies du trésor de Juillac : Jean-François Peiré (DRAC\footnote{Direction Régionale des Affaires Culturelles (DRAC)} Midi-Pyrénées).
\end{itemize}


\section*{Liste des acronymes}
\addcontentsline{toc}{section}{Liste des acronymes}  

\begin{description}
  \item[ACCADIL] \textit{Ancient Coin Classification Algorithm for DIe Links}
  \item[APN] Appareils Photos Numériques
  \item[CNRS] Centre National de la Recherche Scientifique
  \item[CPU] Central Processing Unit
  \item[DRAC] Direction Régionale des Affaires Culturelles
  \item[HALMA] Histoire Archéologie Littérature des Mondes Anciens
  \item[IHM] Interface Homme-Machine
  \item[PDF] Portable Document Format
  \item[RHAdAMANTE] Recherches en Histoire et Archéologie des Âges des Métaux et de l'Antiquité en Europe
  \item[TRACES] Travaux et Recherches Archéologiques sur les Cultures, les Espaces et les Sociétés
\end{description}

\end{document}